\documentclass[preprint2]{aastex}
\usepackage{rotating}
\shortauthors{Nu\~nez et al.}

\begin{document}


\title{Symbiotic stars in X-rays III: long term variability.}


%\author{Natalia, Tommy, Koji, Jeno, Juan}

\author{N. E. Nu\~nez\altaffilmark{1}, T. Nelson\altaffilmark{2}, K. Mukai\altaffilmark{3}, J. L. Sokoloski\altaffilmark{4} and G. J. M. Luna\altaffilmark{5} }

\altaffiltext{1}{Instituto de Ciencias Astron\'omicas de la Tierra y del Espacio (ICATE-UNSJ, CONICET), Av. Espa\~na (S) 1512, J5402DSP, San Juan, Argentina}
\email{nnunez@icate-conicet.gov.ar}

\altaffiltext{2}{Minnesota Institute for Astrophysics, University of Minnesota, Minneapolis, MN, 55455, USA}
\altaffiltext{3}{CRESST and X-ray Astrophysics Laboratory, (NASA/GSFC), Greenbelt MD 20 771, USA. Department of Physics, University of Maryland, Baltimore County, 1000 Hilltop Circle, Baltimore, MD, 21 250, USA)}
\altaffiltext{4}{Columbia Astrophysics Lab, 550 W120th St., 1027 Pupin Hall, MC 5247 Columbia University, 10027, New York, USA}
\altaffiltext{5}{Instituto de Astronom\'ia y F\'isica del Espacio (IAFE, CONICET-UBA),
	Av. Inte. G\"uiraldes 2620, C1428ZAA, Buenos Aires, Argentina}

%*****************************


\begin{abstract}

% * <jeno@astro.columbia.edu> 2015-04-20T20:08:32.933Z:
%
%  I think it would be great if we could add a sentence motivating this research to the beginning of the abstract.
%
We study the X-ray emission from five symbiotic stars observed with {\em Suzaku}. These objects were selected for deeper observations with {\em Suzaku} after their first detection with ROSAT and {\em Swift}. We found that the X-ray spectra can be adequately fit with absorbed optically thin thermal plasma models, either single or multi-temperature. Such a model is compatible with the X-ray emission being originated in the innermost region of the accretion disk, i.e. a boundary layer. 
Based on the large flickering amplitude (only detected in 4~Dra), the high plasma temperature and previous measurements of UV variability and luminosity, we conclude that all five sources are accretion-powered through predominantly opticall thick boundary layer. 

Given the time lapse between previous and these observations, we were able to study the long term variability of their X-ray emission and found that the intrinsic X-ray flux and the intervening absorption column can vary by factors of three or more. However, it is still elusive the location of the absorber and how the changes in the accretion rate and absorption are related. 
%likely due to changes in the accretion rate.  

%. In particular, we found that the X-ray spectrum of Hen~3-1591 is compatible with the presence of a white dwarf instead of a neutron star as accretor. {\bf abstract needs improvements....}

\end{abstract}



\section{Introduction}

When observed at optical wavelengths, symbiotic stars (SS) show a composite spectrum which suggests that they are binary systems. A hot, compact component (usually a white dwarf, WD) contributes to the blue-UV region of the spectrum, while a cool red giant dominates the spectrum at longer wavelengths.  Observations at other wavelengths reveal a very complex and rich scenario for these systems.  Optical, infrared and UV spectral regions are rich in emission lines from forbidden and permitted transitions, which arise mainly from photoionization and recombination of the nebular plasma heated by the the hot component \citep{kenyon09}. Radio observations reveal jets \citep{brocksopp04, crocker01, kellogg07} and thermal emission from the ionized red-giant wind  \citep{seaquist90, seaquist93}. Symbiotic stars can even produce $\gamma$-rays during nova-type outbursts \citep[e.g. V407 Cyg; V745 Sco][]{v407cyg}. 

Symbiotics are now recognized as a population of X-ray sources. From the $\sim$ 220 systems known, 45 have been detected at X-ray wavelengths, most of them with emission in the 0.3 to 10 keV range. A few, however, were detected at energies up to 100 keV \citep{kennea09}. The X-ray spectra from symbiotic stars can be divided into five categories (\cite{Paper1} hereafter Paper I), according to the nature of their emission (thermal or non-thermal) and the maximum energy of their spectra. \citet{Paper1} and \citet{nunez14} studied and classified the first X-ray detections of many symbiotics with {\em Swift}, XMM-{\em Newton} and/or {\em Chandra}. They found that thermal X-ray emission (optically thin or blackbody-type) arises from the accretion-disk boundary layer, a colliding-wind region or quasi-stable nuclear burning on the WD surface.
% * <jeno@astro.columbia.edu> 2015-04-20T20:46:01.677Z:
%
%  I think it might be more useful to the reader if we describe the X-ray emission as consisting of some combination of 4 distinct spectral components (alpha, beta, gamma, delta), which most likely arise from distinct emission regions and/or processes.
%
% ^ <jeno@astro.columbia.edu> 2015-04-20T20:47:26.196Z.

Sensitive, broadband X-ray satellites such as {\em Suzaku} have also played a significant role in observing symbiotics, especially those with X-ray emission above 10 keV \citep[T CrB, CH Cyg, V648 Car;][]{luna08, mukai07, kennea09}. Here we present the analysis of {\em Suzaku}/XIS observations of a set of five symbiotics previously known to be X-ray sources.
%[and selected for deeper observations with {\em Suzaku}]. 
This sample consist of \object{CD-28~3719}, \object{EG~And}, \object{Hen~3-461}, \object{Hen~3-1591} and \object{4 Dra}. In Section \ref{sec:obs-red} we detail the data reduction and analysis, while in Section \ref{sec:results} we present the results. Concluding remarks are presented in Section \ref{sec:concl}. 
% * <jeno@astro.columbia.edu> 2015-04-20T20:59:38.568Z:
%
%  I would love it if after our telecon, we change this final sentence to something like, "In Section xx, we explain our conclusion that yy," where yy is some nice finding that we want readers to take away from this paper.
%
% ^ <jeno@astro.columbia.edu> 2015-04-20T21:01:51.686Z.

\section{Observations and data reduction}
\label{sec:obs-red}

The X-ray Imaging Spectrometer (XIS) on board {\it Suzaku} observed five symbiotic stars. Details of each observation are presented in Table \ref{table:info1}. All sources were observed with the XIS0, XIS1, and XIS3 detectors (the XIS2 is not operative since November 2006 \citep{koyama07}, which are sensitive in the 0.2-12.0 keV range. All sources were too faint to be detected with the Hard X-ray detector (HXD).  We reprocessed all data using the \texttt{aepipeline} script and obtained event files with the processing version 2.5.16.29 (2014-07-01) applied.  

Source spectra and light curves were extracted from circular regions centered on the source SIMBAD\footnote{http://simbad.u-strasbg.fr/simbad/sim-fid} coordinates. The recommended radius of the extraction region\footnote{\url{http://heasarc.gsfc.nasa.gov/docs/suzaku/analysis/abc/}} is 260$\arcsec$ (this size encircles 99\% of the point source flux), however given the source and background brightness, we were able to use this size only in the case of 4~Dra. Comparing the source spectrum with that from the background, we found that the optimal radius for the source region, which maximizes the signal-to-noise ratio, was of 120$\arcsec$ in the case of Hen~3-1591, CD~-28~3719 and EG~And and 60$\arcsec$ for Hen~3-461. Background spectra and light curves were extracted from annulus regions centered on the source (with inner and outer radius of 200$\arcsec$ and 400$\arcsec$, respectively) in the case of Hen~3-1591, CD~-28~3719, Hen~3-461 and 4~Dra while a circular region with 160$\arcsec$ radius was used for EG~And 
because the location of the source on the chip did not allowed to select an annulus region for the background. The response matrices were created using the \texttt{rmfgen} and \texttt{arfgen} scripts. We then fit the binned spectra (grouping by a minimum of 16 to 20 counts per bin) using XSPEC\footnote{http://heasarc.gsfc.nasa.gov/docs/xanadu/xspec/} and $\chi^{2}$ as statistic to select the best fit model.

Hen 3-1591 was observed serendipitously with ASCA \citep{tanaka94} on 1999 September 22 during an observation of the supernova remnant (SNR), \object{G5.2-2.6}. The Solid-state Imaging Spectrometer (SIS) was operated in 1-CCD mode for this observation (of the 4 chips available), which put Hen 3-1591 outside the operational SIS field of view. On the other hand, Hen 3-1591 was securely in the field of view of the Gas Imaging Spectrometer (GIS) instrument, which has two units (GIS2 and GIS3). We have selected intervals when the satellite was outside the South Atlantic Anomaly (SAA), the attitude control was stable and within 0.02 degrees of the target, and the line of sight was greater than 5 degrees above the Earth limb. We have also applied a standard selection expression combining monitor count rates and geomagnetic cutoff rigidity (COR) to exclude time intervals of high-particle background, obtaining 15 ks of good on-source data.

For the ASCA observation on Hen~3-1591 we have extracted the source spectrum from a 6$^{\prime}$ radius circular extraction region centered on the source, and background spectrum from a 6$^{\prime}$ radius region centered at (271.6619, -25.8192) away from the SNR and other obvious sources. Hen~3-1591 is detected at a net count rate of 0.016 c/s/GIS. We use the v4.0 rmf downloaded from the CALDB and have generated an arf file for each unit of the GIS. We have then combined the GIS2 and GIS3 spectra and responses, ignored the data outside the well-calibrated range of 0.7--10 keV, and binned the data by a factor of 32, leaving 26 channels.

%is located in a crowded region on the Galactic plane, including the presumed X-ray emission from the SNR. 

%Hen~3-461 was in a very faint state during the observation to allow us to extract products (see Section \ref{sec:reshen3461}). 

\begin{deluxetable}{l r c c}
%\rotate
\tabletypesize{\footnotesize}
\tablecolumns{4}
\tablewidth{0pt}
\tablecaption{Observations Log \label{table:info1}}
\tablehead{
\colhead{Source} & \colhead{Date} & \colhead{ObsId}  & \colhead{Exp. Time [ks]}}   % \colhead{Coordinates (J2000)} & \colhead{Search-Offset [arcmin]}} 
\startdata
\hbox to 0.9in{4~Dra\leaders\hbox to 0.8em{\hss.\hss}\hfill} & 2010-04-18 & 405035010 & 42 \\ %& 12 30 06.66 +69 12 04.06 &  & 1.029 \\
                                                                        & 2011-11-09 & 406041010 & 42 \\ % 12 30 06.66 +69 12 04.06 &  & 1.201 \\
\hbox to 0.9in{CD~-28~3719\leaders\hbox to 0.8em{\hss.\hss}\hfill}   & 2013-10-12 & 408032010 & 14 \\ % 07 01 08.78 -29 07 00.5  &  & 0.597 \\
\hbox to 0.9in{EG~And\leaders\hbox to 0.8em{\hss.\hss}\hfill}        & 2011-02-05 & 405034010 & 100  \\ % 00 44 40.32 +40 40 22.8 & & 0.708\\
\hbox to 0.9in{Hen~3-461\leaders\hbox to 0.8em{\hss.\hss}\hfill}      & 2012-12-17 & 407007010 & 46 \\ % 10 39 06.70 -51 24 31.7  &  & 0.447  \\ 
\hbox to 0.9in{Hen~3-1591\leaders\hbox to 0.8em{\hss.\hss}\hfill}     & 2012-10-03 & 407042010 & 51  \\ % 18 07 32.71 -25 53 53.2  & & 0.226\\
\hbox to 0.9in{- ASCA \leaders\hbox to 0.8em{\hss.\hss}\hfill} & 1999-09-22 & 57055000  & 15 \\
\enddata
\end{deluxetable}

We evaluate whether any of the three spectral models, i.e. an absorbed single-temperature optically thin thermal plasma, an absorbed multi-temperature cooling flow or an absorbed non-thermal plasma properly fits the data by their $\chi^{2}_{\nu}$. Those best-fit models are described in the following sections while the resulting parameters are listed in Table \ref{tab:models}.  With the selected model, we use the unbinned data and C-statistic \cite{cash} to calculate the uncertainties in the parameters of the model and the flux. All errors in the fit parameters were estimated at their 90\% confidence (see Table \ref{tab:models}).

\section{Results}
\label{sec:results}

\subsection{Spectral analysis}

The five sources observed with {\em Suzaku} were modeled with absorbed optically thin thermal emission, which in the case of \object{Hen~3-1591}, suggest that the compact object is a white dwarf instead of a neutron star as previously thought. All sources show strong changes in their X-ray flux and the hardness in long time scales, from a few months to years, which could be related with changes in the accretion rate and/or absorption.  

\subsubsection{4~Dra}% (=CQ~Dra)}
\label{sec:res4dra}

This source was classified as a triple system consisting of 4~Dra(A) + CQ~Dra(Bab) \citep{reimers85}. However, X-ray data obtained with ROSAT\footnote{http://www.xray.mpe.mpg.de/cgi-bin/rosat/rosat-survey} lead \citet{wheatley03} to suggest that the X-ray emission is consistent with the presence of a WD accreting from the wind of the red giant and the UV data from IUE led \citet{skopal} to the same conclusion after fitting the SED. 
%It has the characteristics of a symbiotic binary with a M3~II giant component and a hot companion, probably cataclysmic or maybe an AM~Her system (CQ~Dra) \citep{reimers88}. 
More evidence against the triple system nature of 4~Dra comes from the analysis of the broad wings superimposed to narrow emission lines in FUSE spectra by \citep{froning12}, similar to other FUSE spectra of confirmed symbiotic stars.
%Froning et al. (2012ApJS..199....7F)
This source is one of the two system with Hipparcos distance in our sample, $d$=190$\pm$17 pc \citep{vanleuween07}. Its optical brightness presents irregular variations about 0.1 mag in V \citep{eggen67} and have radio flux variations on time scales of weeks to months and maybe shorter than hours to days \citep{brown87}. 
%After four observations (including the adquired with ROSAT, %\citep{trumper83} we found strong variability on time scales of minutes to years.
%, leading \citet{wheatley03} to suggest that the X-ray emission is consistent with the presence of an accreting WD from the wind of the giant. They did not find periodic variations in X-rays, expected if the hot companion was part of a AM~Her system as proposed by \citet{reimers88}. 
Based on these previous studies, in order to model the {\em Suzaku} spectra, we consider this source to be a symbiotic system with an accreting white dwarf.

We analyzed the two X-ray observations, ObsID 405035010 and 406041010 (hereafter 4050 and 4060, respectively. See Table \ref{table:info1}). We obtained acceptable fits with two spectral models in two different scenarios. 
In the first scenario, we forced the temperature of the optically thin thermal emission to have a unique value for both observations. This is a valid assumption if the X-ray emitting plasma arise in the post-shock region of the accretion disk boundary layer and its temperature is set by the WD mass, which should not change between observations. In the second scenario, we modeled both observations independently. 

%READY
The first spectral model consists on a single temperature APEC\footnote{https://heasarc.gsfc.nasa.gov/xanadu/xspec/manual/XSmodelApec.html} plasma model with a reduced metalicity, observed through a simple and partial covering absorber. For the first scenario the plasma temperature and abundances are $kT$=6.0$\pm$0.3 keV and abundance=0.18$\pm$0.02; the absorption column of the simple absorber is $N_{H,22}^{4050}$=0.43$\pm$0.03 and $N_{H,22}^{4060}$=0.94$\pm$0.06; while for the partial covering absorber we have $N_{H,22}^{4050}$=1.26$\pm$0.5, $N_{H,22}^{4060}$=2.61$\pm$0.22; covering fractions of CF(4050)=0.43$\pm$0.03 and CF(4060)=0.64$\pm$0.03; unabsorbed fluxes F$_{X}^{4050}$=136$\pm$2$\times$10$^{-13}$ ergs cm$^{-2}$ s$^{-1}$, F$_{X}^{4060}$=697$\pm$2$\times$10$^{-13}$ ergs cm$^{-2}$ s$^{-1}$ and $\chi^{2}_{\nu}$=1.06.

%READY
If we use the model above to fit both observations independently, we obtain $kT$(4050)=4.5$\pm$0.1 keV and $kT$(4060)=6.4$\pm$0.1 keV; 
abundances of 0.17 (4050) and 0.19 (4060); $N_{H,22}^{4050}$=0.46$\pm$0.03 and $N_{H,22}^{4060}$=0.88$\pm$0.07; while for the partial-covering absorber we have $N_{H,22}^{4050}$=2.5$\pm$0.5, $N_{H,22}^{4060}$=2.3$\pm$0.2; covering fractions of CF(4050)=0.38$\pm$0.04 and CF(4060)=0.66$\pm$0.04; unabsorbed fluxes F$_{X}^{4050}$=136$\pm$2$\times$10$^{-13}$ ergs cm$^{-2}$ s$^{-1}$, F$_{X}^{4060}$=564$\pm$2$\times$10$^{-13}$ ergs cm$^{-2}$ s$^{-1}$ and $\chi^{2}_{\nu}$(4050)=1.02, $\chi^{2}_{\nu}$(4060)=1.07. 

% READy
As an alternative, the second model consists on a multi-temperature cooling flow model observed through a simple and partial covering absorber. In the case that the X-ray emission from both observations originates from a plasma with the same temperature and abundances, we obtain $kT$=10.8$_{-0.5}^{+0.6}$ keV and abundance=0.21$\pm$0.01; the absorption column of the simple absorber is $N_{H,22}^{4050}$=0.54$\pm$0.07 and $N_{H,22}^{4060}$=1.21$\pm$0.06; while for the partial-covering absorber we have $N_{H,22}^{4050}$=1.5$\pm$0.5, $N_{H,22}^{4060}$=3.5$\pm$0.3; covering fractions of CF(4050)=0.39$_{-0.09}^{+0.11}$ and CF(4060)=0.60$\pm$0.03; unabsorbed fluxes F$_{X}^{4050}$=145$\pm$2$\times$10$^{-13}$ ergs cm$^{-2}$ s$^{-1}$, F$_{X}^{4060}$=723$\pm$2$\times$10$^{-13}$ ergs cm$^{-2}$ s$^{-1}$ and $\chi^{2}_{\nu}$=1.05.

% READY
If we model both observations independently with an absorbed cooling flow model, we obtain $kT$(4050)=7.3$_{-0.8}^{+1.0}$ keV and $kT$(4060)=12.9$_{-0.8}^{+0.9}$ keV; 
abundances 0.18$\pm$0.03 (4050) and 0.22$\pm$0.02 (4060); $N_{H,22}^{4050}$=0.64$\pm$0.05 and $N_{H,22}^{4060}$=1.13$\pm$0.07; while for the partial-covering absorber we have $N_{H,22}^{4050}$=3.2$\pm$0.9, $N_{H,22}^{4060}$=2.8$\pm$0.3; covering fractions of CF(4050)=0.42$\pm$0.04 and CF(4060)=0.61$\pm$0.04; unabsorbed fluxes F$_{X}^{4050}$=175$\pm$2$\times$10$^{-13}$ ergs cm$^{-2}$ s$^{-1}$, F$_{X}^{4060}$=673$\pm$2$\times$10$^{-13}$ ergs cm$^{-2}$ s$^{-1}$ and $\chi^{2}_{\nu}$(4050)=1.01 $\chi^{2}_{\nu}$(4060)=1.07.

Although in terms of $\chi^{2}_{\nu}$ an absorbed non-thermal plasma plus a gaussian emission line (\texttt{wabs$\times$(power+gauss)}) fit the observed spectrum, the line centroid is at $\sim$6.67 keV, consistent with \ion{Fe}{25} transitions from a thermal plasma. We thus opt for a thermal origin for the observed X-ray emission.

\subsubsection{CD~-28~3719}

\label{sec:rescd-28}

In X-rays, \object{CD~-28~3719} was detected for the first time in a short pointing observation with {\em Swift} and its spectrum classified as $\delta$-type, i.e. optically thin thermal, hard X-ray emission observed through strong absorption (Paper I).  Our {\em Suzaku} observation aimed to obtain a spectrum with a higher signal-to-noise ratio in order to improve the basic spectral modeling. During the observation there were some problems in the acquisition of the XIS0 chip data which did not go back to the 5$\times$5 editing mode after dark frame dump during a South Atlantic Anomaly passage and the data during these segments were corrupted. For this reason, we only analyzed the data of the XIS1 and XIS3 chips. 

The best fit model for this source consist on an absorbed cooling flow model with variable abundance \texttt{wabs$\times$mkcflow}. The absorption column $N_{H,22}$ is high, 12.9$^{+3.4}_{-2.8}$, the maximum temperature $kT_{max}$ is 8$_{-3}^{+6}$ keV and the abundance is 0.44$^{+0.30}_{-0.17}$. The unabsorbed flux (in the 0.3-10 keV range) F$_X$ is 23$^{+2}_{-2}\times$10$^{-13}$ erg cm$^{-2}$ s$^{-1}$. The luminosity L$_X$, at a distance of 1 kpc, is 28$^{+2}_{-2}\times$10$^{31}$ ergs s$^{-1}$ (d/1 kpc)$^{2}$. The results from this fit are commensurate with the results obtained from the {\em Swift} data analyzed in Paper I.

\subsubsection{EG And}
\label{sec:reseg}

In the V band, EG~And is one of the brightest symbiotic systems. Its distance is 512$\pm$168 pc \citep{vanleuween07}. The periodic photometric modulation indicates an orbital period of 470 days with an inclination of 82$^{+8}_{-4.5}$ degrees \citep{kolb04}, making EG~And an eclipsing symbiotic binary. Our {\em Suzaku} observation took place at the orbital phase $\phi$=0.93 \citep[using the ephemeris from][]{kolb04}, with the WD going behind the red giant wind or 
%{\bf please double check this ephemeris} 
$\phi$=0.17 is we use the orbital period of 481 days and the ephemeris from Vogel et al. (1991), i.e. the WD coming out of the partial eclipse.
%{\bf If we use Vogel 1991, of Belckzynski to ephemeris:Min(UV ) = JD 2445380 + 481E, with JD$_{obs}$=2455598.128, we obtain phase=21.17 then 0.17}

%and most of them with energies $\gtrsim$ 1 keV 

After visual inspection, we verified that no obvious features were present in the spectrum and group the channels to have a minimum of 20 counts per bin, which allow us to use the $\chi^{2}$ statistics during the fit. The best fit model consists of an absorbed single plasma temperature \texttt{wabs$\times$apec}, with absorption column $N_{H,22}$=0.09$_{-0.07}^{+0.09}$ and temperature $kT$= 7$_{-2}^{+3}$ keV. The unabsorbed flux is F$_X$=1.3$\pm$0.2$\times$10$^{-13}$ ergs s$^{-1}$ cm$^{-2}$ and the luminosity L$_{X}$=0.4$\times$10$^{31}$ ergs s$^{-1}$.

%(d/512 pc)$^{2}$. 

%{\bf We cannot estimate the M$_{WD}$ taking into account the relation between the k$_{max}$ of the cooling flow model because the temperature is so low (why?). May be another cooling mechanism is acting first to reduce the shock temperature?.}

%Using the relation between  N$_{H}$ and (B-V) proposed by Groenewegen \& Lamers 1989, we calculated a value of 1.53 mag. This value is bigger than 0.05 mag proposed Crowley 2008. Maybe the relation needed some recalibration...

\subsubsection{Hen 3-461}
\label{sec:reshen3461}

Hen~3-461 was discovered in X-rays with {\em Swift} during a short pointing observation (Paper I).  The high temperature and absorption obtained from modeling those data, the hardness ratio (defined as the ratio of count rates in the 2.4-10/0.3-10 keV ranges) and the presence of significant flickering in the UV, suggested that the X-ray emission originates in the boundary layer of the accretion disk and lead us the authors classify it as a $\delta$-type source. 
Almost two and half years later, {\em Suzaku} observed Hen~3-461 and only detected 285 source photons, i.e. 8$\sigma$ detection after background substraction.  
Given the low number of photons detected, we performed a crude spectral modeling ($\chi^{2}_{\nu}$=1.45). We noticed that a weak soft component is detected at a low significance level (2$\sigma$) and thus we decided to include a second component in our spectral model, being aware of the similarity with the well-known, two-spectral components $\beta/\delta$-type emission observed in few white dwarf symbiotics (e.g. NQ~Gem, V347~Nor, see Paper I). Our best spectral model consist of two optically thin thermal components ($kT_{1}$=0.07$\pm$+0.02 keV and $kT_{2}$=8$_{-3}^{+5}$ keV) and a gaussian emission line at 6.4 keV. The hardest component is modified by a full ($N_{H,22}$=3$_{-2}^{+5}$) and a partial covering absorber ($N_{H,22}^{pc}$=75$\pm$25: covering fraction=0.97$\pm$0.02). The unabsorbed X-ray flux is F$_X$=26$\pm$5$\times$10$^{-13}$ ergs s$^{-1}$ cm$^{-2}$ and assuming a distance of 1 kpc, the luminosity is L$_{X}$=31$\pm$1$\times$10$^{31}$ ergs s$^{-1}$.

%, which precludes any meaningful spectral fit. 

\subsubsection{Hen~3-1591}

\label{sec:reshen315}

Only two objects were classified by \citet{muerset97} as $\gamma$-type symbiotics based on their emission detected with ROSAT: GX~1+4 and Hen~3-1591. These observations had limited signal-to-noise ratio but \citet{muerset97} speculated that these sources harbored neutron stars as accretors. Lately, these type of symbiotics were named as ''symbiotic X-ray binaries" by \citet{masetti07}. 

\begin{figure}
 \includegraphics[scale=0.3]{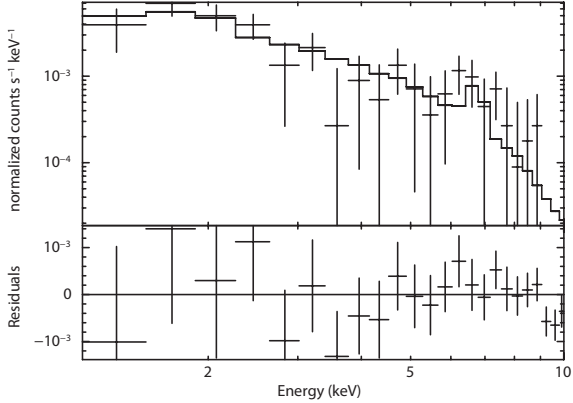}
 \caption{{\em ASCA}/GIS spectra of Hen~3-1591. The full line shows the best-fit model described in Section \ref{sec:results}. } 
\label{fig:asca}
\end{figure}

We have found that Hen~3-1591 was serendipitously observed with ASCA/GIS (but outside the SIS field of view) in 1999, with an exposure time of 16 ksec. The spectrum is not of high enough statistical quality (see Fig. \ref{fig:asca}) to draw definitive conclusions (only $\sim$200 photons were detected), but some inferences can be drawn.  A power-law fit prefers a photon index near 1.9, with excess counts around 6.5 keV. Adding a Gaussian, the line centroid was found to be near 6.6 keV with an equivalent width well in excess of 1 keV. We interpret this as due to a combination of Fe H-like, He-like, and fluorescent lines, i.e., the X-ray emission should be modeled as optically thin thermal emission with reflection adding the 6.4 keV line.  In fact, a cooling flow plus Gaussian model fit gives a good description of the observed spectrum.  The maximum temperature is around 14 keV, iron is strongly overabundant ($\sim$ twice the  solar value), and the equivalent width of the fluorescent line ($\sim$ 600 eV) also 
requires an overabundance of iron. Taken at face value, this suggests that Hen~3-1591 hosts a white dwarf accreting matter overabundant in Fe. Our {\em Suzaku} spectrum aimed to identify, in a higher signal-to-noise spectrum, the best spectral model. Given that ROSAT data suggested a non-thermal origin for the X-ray emission while ASCA data suggested that the X-ray emission is due to a multi-temperature optically thin thermal plasma, we try these two models in the {\em Suzaku} spectrum. First we try two thermal models, which show the same statistic ($\chi^{2}_{\nu}$=1.07) and similar model parameters. We prefer the multi-temperature model, \texttt{wabs$\times$mkcflow}, which yielded an absorption column of $N_{H,22}$=0.11$_{-0.06}^{+0.05}$ and a maximum temperature  $kT_{max}$=7$_{-2}^{+3}$ keV. %with this value we can obtain a M$_{WD}$= 0.45 M$_{\odot}$ and the R$_{WD}$= 10.3$\times$10$^{8}$ cm.
% This line should go in the discusion together with the others
The unabsorbed flux is F$_X$=7.4$\pm$0.3$\times$10$^{-13}$ ergs cm$^{-2}$ s$^{-1}$, and L$_X$ at a distance of 1 kpc is 8.9$\pm$0.4$\times$10$^{31}$ ergs s$^{-1}$.

%his model has a $\chi^{2}_{\nu}$=1.07 and it is consistent with the assumption of the accretor being a WD and not a neutron star.

%The Gaussian component was centered near 6.6 keV with an equivalent width of 0.84 keV. The goodness parameter for this model is 70\% and $\chi^{2}_{\nu}$=1.11 (190 degrees of freedom), and goodness is 70\%. 

%As a second model, consistent with the presence of an accreting neutron star, we used an absorbed Comptonized spectrum of cool photons on hot electrons, \texttt{wabs$\times$(compST+gauss)} \citep{sunyaev80} which resulted in an absorption column of $N_{H,22}$=0.20$_{-0.05}^{+0.06}$, an electron temperature kT$_{compST}$=9.68$_{-0.67}^{+0.75}$ keV, optical depth $\tau$=4.74$_{-0.46}^{+0.48}$ and $\chi^{2}_{\nu}$=1.14. The unabsorbed X-ray flux is F$_X$=10.5$\pm$0.5$\times$10$^{-13}$ ergs cm${^-2}$ s$^{-1}$, and the luminosity L$_{X}$=12.6$\pm$0.6$\times$10$^{31}$ ergs s$^{-1}$ (d/1 kpc)$^{2}$. 

On the other hand, an absorbed non-thermal model, 
%We can consider a more simple model to evaluate this possibility, 
\texttt{wabs$\times$powerlaw} yields $N_{H,22}$=0.21$\pm$0.06 and the photon index is 2.40$\pm$0.19. The absorbed flux F$_X$=10.7$\pm$0.5$\times$10$^{-13}$ ergs cm$^{-2}$ s$^{-1}$, and the luminosity L$_{X}$=12.9$\pm$0.6$\times$10$^{31}$ ergs s$^{-1}$ (d/1 kpc)$^{2}$, $\chi^{2}_{\nu}$=1.11.

%The $\chi^{2}_{\nu}$=1.14 with 174 d.o.f., this is a good fit but not the best.
%The goodness parameters for this fits was 64\%, with $\chi^{2}_{\nu}$=1.08 

\begin{figure*}
 \includegraphics[]{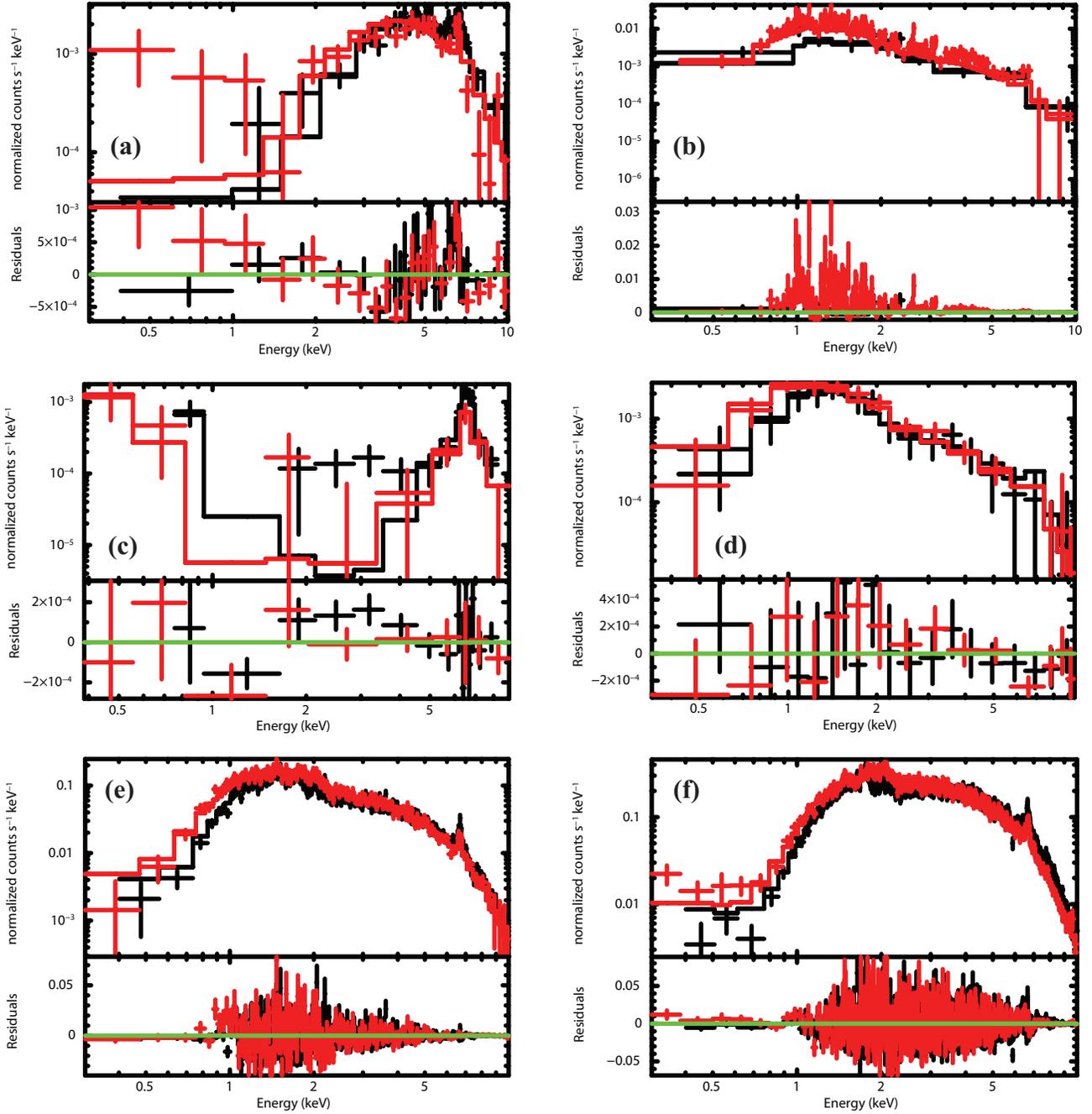}
 \caption{{\em Suzaku}/XIS spectra of ($a$) CD~-28~3719, ($b$) Hen~3-1591, ($c$) Hen 3-461, ($d$) EG~And and ($e$) 4~Dra ObsID 4050 and ($f$) 4~Dra ObsID 4060. The full line shows the best-fit model described in Section \ref{sec:results}. Red (XIS1), black (XIS0+3)} 
\label{fig1}
\end{figure*}

%\rotate
\renewcommand{\arraystretch}{1.5}
\begin{deluxetable}{l l c c c c c c}
%\rotate
%\begin{sidewaytable}{h}
\tabletypesize{\scriptsize}
\tablecolumns{9}
\tablewidth{0pt}
\tablecaption{Spectral Models \label{tab:models}}

\tablehead{\colhead{Object} & \colhead{Model} & \colhead{Count rate} & \colhead{N$_{H,22}$} & \colhead{kT\tablenotemark{a}}  & \colhead{F$_{X}$\tablenotemark{b}} & \colhead{L$_{X}$\tablenotemark{c}} & \colhead{$\chi^{2}_{\nu}$} \\
 & & [10$^{-2}$ counts s$^{-1}$] &  &  [keV] &  & &\\ 
 }
\startdata
%\hline
1st scenario &&&&&& \\
(see Sect. \ref{sec:res4dra}) &&&&&& \\
\hline
\hbox to 0.8in{\object{{4~Dra}(4050)}   \leaders\hbox to 0.4em{\hss.\hss}\hfill}   &  \texttt{wabs$\times$pcfabs$\times$vapec} & 38$\pm$1 & 0.43$\pm$0.03 & 6.0$\pm$0.3  & 136$\pm$2 & 6$\pm$1 & 1.06 \\
 &  &  & 1.26$\pm$0.50, CF=0.43$\pm$0.03 & &  &  &  \\

\hbox to 0.8in{\object{{4~Dra}(4060)}   \leaders\hbox to 0.4em{\hss.\hss}\hfill}&  \texttt{wabs$\times$pcfabs$\times$vapec} &110$\pm$1 & 0.94$\pm$0.06 & 6.0$\pm$0.3 &  697$\pm$2 & 30$\pm$1 & 1.06\\
 &  &  & 2.61$\pm$0.22, CF=0.64$\pm$0.03 & &  &  &  \\

\hbox to 0.8in{\object{{4~Dra}(4050)}   \leaders\hbox to 0.4em{\hss.\hss}\hfill}&  \texttt{wabs$\times$pcfabs$\times$mkcflow} & $\cdots$ & 0.54$\pm$0.07 & 10.8$_{-0.5}^{+0.6}$ & 145$\pm$2 & 6$\pm$1 & 1.05 \\
 &  &  & 1.5$\pm$0.5, CF=0.39$_{-0.09}^{+0.11}$ & &  &  &  \\

\hbox to 0.8in{\object{{4~Dra}(4060)}   \leaders\hbox to 0.4em{\hss.\hss}\hfill}&  \texttt{wabs$\times$pcfabs$\times$mkcflow} & $\cdots$ & 1.21$\pm$0.06 & 10.8$_{-0.5}^{+0.6}$   & 723$\pm$2 & 31$\pm$1 & 1.05 \\
 &  &  & 3.5$\pm$0.3, CF=0.60$\pm$0.03 & &  &  &  \\
%\\
\hline
2nd scenario &&&&&& \\
(see Sect. \ref{sec:res4dra}) &&&&&& \\
\hline
\hbox to 0.7in{\object{{4~Dra}(4050)}   \leaders\hbox to 0.4em{\hss.\hss}\hfill} &  \texttt{wabs$\times$pcfabs$\times$vapec} & $\cdots$ & 0.46$\pm$0.03 & 4.5$\pm$0.1  & 136$\pm$2 & 6$\pm$1 & 1.02 \\
 &  &  & 2.5$\pm$0.5, CF=0.38$\pm$0.04 & &  &  &  \\

\hbox to 0.7in{\object{{4~Dra}(4060)}   \leaders\hbox to 0.4em{\hss.\hss}\hfill}&  \texttt{wabs$\times$pcfabs$\times$vapec} & $\cdots$ & 0.88$\pm$0.07 & 6.4$\pm$0.1 &  564$\pm$2 & 24$\pm$1 & 1.07\\
 &  &  & 2.3$\pm$0.2, CF=0.66$\pm$0.04 & &  &  &  \\

\hbox to 0.8in{\object{{4~Dra}(4050)}   \leaders\hbox to 0.4em{\hss.\hss}\hfill}&  \texttt{wabs$\times$pcfabs$\times$mkcflow} & $\cdots$ & 0.64$\pm$0.05 & 7$\pm1$ & 175$\pm$2 & 8$\pm$1 & 1.01 \\
 &  &  & 3.2$\pm$0.9, CF=0.42$\pm$0.04 & &  &  &  \\

\hbox to 0.8in{\object{{4~Dra}(4060)}   \leaders\hbox to 0.4em{\hss.\hss}\hfill}&  \texttt{wabs$\times$pcfabs$\times$mkcflow} & $\cdots$ & 1.13$\pm$0.07 & 12$\pm$1 & 673$\pm$2 & 29$\pm$1 & 1.07 \\
 &  &  & 2.8$\pm$0.3, CF=0.61$\pm$0.04 & &  &  &  \\
\hline
\hbox to 0.8in{\object{CD~-28~3719}   \leaders\hbox to 0.4em{\hss.\hss}\hfill} &  \texttt{wabs$\times$apec} & 0.9$\pm$0.1  & 12$_{-4}^{+2}$  & 5$\pm$2 &  17$\pm$1 & 20$\pm$1 & 1.17 \\
			 &  \texttt{wabs$\times$mkcflow} & $\cdots$ & 13$\pm$3 & 8$_{-3}^{+6}$ & 23$\pm$1 & 28$\pm$2 & 1.21 \\
%			 &  \texttt{wabs$\times$powerlaw} & $\cdots$ & 15.73$_{-4.05}^{+5.46}$  & $\cdots$ & 3.09$_{-0.65}^{+0.80}$  & 126.0$^{+9.0}_{-9.0}$ & 15.1$^{+1.0}_{-3.4}$ & 1.13 \\
\\
\hbox to 0.8in{\object{EG And}   \leaders\hbox to 0.4em{\hss.\hss}\hfill} &  \texttt{wabs$\times$apec} & 0.4$\pm$0.1 & 0.09$_{-0.07}^{+0.09}$ & 7$_{-2}^{+3}$ &  1.3$\pm$0.2 & 0.4$\pm$0.1 & 0.9 \\
%\object{EG And} &  \texttt{wabs$\times$mkcflow} & $\cdots$ & 1.54$_{-0.27}^{+0.30}$ & 0.72$_{-0.30}^{+0.32}$  & $\cdots$ & 11.0$^{+2.2}_{-1.9}$ & 3.5$^{+0.7}_{-0.6}$ & 0.91 \\
%\object{EG And} &  \texttt{wabs$\times$powerlaw} & $\cdots$  & 0.56$_{-0.19}^{+0.22}$ & $\cdots$  & 3.55$\pm$0.41 & 3.9$^{+0.3}_{-1.1}$ & 1.1$^{+0.2}_{-0.2}$ & 0.91 \\
\\
\hbox to 0.8in{\object{Hen 3-461}   \leaders\hbox to 0.4em{\hss.\hss}\hfill} &  \texttt{apec+wabs$\times$pcfabs$\times$(apec+gauss)} & 0.2$\pm$0.1  & 3$_{-2}^{+5}$  & 8$^{+5}_{-3}$  &  26$\pm$5 & 31$\pm$6 & 1.45 \\
&   & & 75$\pm$25; CF=0.97$\pm$0.02 \tablenotemark{d} & &   & & \\

			% &  \texttt{wabs$\times$mkcflow} & $\cdots$ & 0.11$_{-0.06}^{+0.05}$ & 7.01$_{-1.73}^{+2.89}$  & $\cdots$ & 7.4$^{+0.3}_{-0.3}$ & 8.9$^{+0.4}_{-0.4}$ & 1.07 \\

\hbox to 0.8in{\object{Hen 3-1591}   \leaders\hbox to 0.4em{\hss.\hss}\hfill} &  \texttt{wabs$\times$apec} & 3.7$\pm$0.1  & 0.05$\pm$0.04  & 3.06$^{+0.55}_{-0.48}$  &  6.6$^{+0.3}_{-0.3}$ & 7.9$\pm$0.4 & 1.07 \\
			 &  \texttt{wabs$\times$mkcflow} & $\cdots$ & 0.11$_{-0.06}^{+0.05}$ & 7$_{-2}^{+3}$  &  7.4$\pm$0.3 & 8.9$\pm$0.4 & 1.07 \\
%			 &  \texttt{wabs$\times$powerlaw} & $\cdots$ & 0.21$\pm$0.06  & $\cdots$ & 2.40$\pm$0.19 & 10.7$^{+0.5}_{-0.5}$ & 12.9$^{+0.6}_{-0.6}$ & 1.11 \\
%\object{Hen 3-1591} & \texttt{wabs$\times$compST} & $\cdots$ & 0.20$_{-0.05}^{+0.06}$  & 9.68$_{-0.67}^{+0.75}$  & $\cdots$  & 10.5$^{+0.5}_{-0.5}$ & 12.6$^{+0.6}_{-0.6}$ & 1.14 \\
\enddata 
\tablenotetext{a}{Indicates the value of the maximum temperature in the case of the cooling flow model \texttt{mkcflow}, $kT_{max}$.}
\tablenotetext{b}{Unabsorbed X-ray flux, in units of 10$^{-13}$ erg s$^{-1}$ cm$^{-2}$ in the  0.3-10.0 keV energy range.}
\tablenotetext{c}{Unabsorbed X-ray luminosity, in units of 10$^{31}$ erg s$^{-1}$ in the  0.3-10.0 keV energy range.}
\tablenotetext{d}{Absorption column and covering fraction of the partial absorber model \texttt{pcfabs}.}
%\hline
% \end{tabular}
%\end{sidewaytable}
\end{deluxetable}

\subsection{Timing analysis}
\label{sec:time}

Significant stocasthic variability of the the X-ray flux in short (minutes to hours) time scales is a hallmark of the presence of accretion disk in binary systems. 
%(Paper 1 {\bf and maybe XMM-proposal from Jeno?}). 
Periodic modulation, on the other hand, would indicate that the accretion is channeled by an strong magnetic field (e.g. Z~And, \cite{sokoloski06}). 

We searched for periodicity using the Lomb-Scargle algorithm on the light curves binned at 16 s. No periods were found for the sources in our sample, with a false alarm probability $\gtrsim$0.3.  We used the ratio between the observed ($s$) and expected ($s_{exp}$) variances to account for stocasthic variations in the light curves. We binned the lightcurves at integer numbers of the XIS readout time: 16, 160, 1600, and 3680 seconds and calculated the ratio $s/s_{exp}$ (see Table \ref{table:timing}).  Only 4~Dra shows significant variability, with amplitudes as high as 40\%, the strongest flickering is present in hours scales but there is also intense flickering in minutes time scale, consistent with previous determinations \citep{wheatley03}. 

\renewcommand{\arraystretch}{1.5}
\begin{deluxetable}{ccccccc}
\tabletypesize{\footnotesize}
\tablecolumns{7}
\tablewidth{0pt}
\tablecaption{Timing analysis \label{tab:time}\tablenotemark{a} \label{table:timing}}
\tabletypesize{\scriptsize}
\tablehead{
\colhead{Object} & \colhead{Bin size [s]} & \colhead{s$_{exp}$} & \colhead{s} & \colhead{Ratio}  & \colhead{s$_{exp}$/average} & \colhead{s/average} 
}
\startdata
%4 Dra& 4050  & 160   & 15.1 & 3.3 \\ %& 81760%
%	&4060  &    & 8.0  & 4.7  \\ %& 76800
%  	&4050  & 1600  & 6.0  & 7.3 \\% & 81760
% 	&4060  &   & 3.2  & 10.3 \\ %& 76800
% 	&4050 & 3680  & 4.4  & 9.0 \\ % & 80960 
% 	&4060  &   & 2.2  & 13.7 \\ %& 73600
% 	&4050  & 7360  & 3.5  & 9.2 \\ %& 80960
% 	&4060  &   & 1.7  & 13.6 \\% & 73600
% 	&4050 & 14720 & 2.5  & 12.3 \\ %& 73600 
% 	&4060  &  & 1.3  & 15.0 \\ %& 73600

%Object & binsize(s) & Sexp & S_measured & ratio_variance & Sensitivity_%_(Sexp/ave_mes) & Sensitivity_%_(Smes/ave_mes)
\hbox to 0.9in{4~Dra(4050)\leaders\hbox to 0.8em{\hss.\hss}\hfill} & 160 & 0.05 & 0.17 & 3.32 & 15 & 50 \\
\hbox to 0.9in{4~Dra(4060)\leaders\hbox to 0.8em{\hss.\hss}\hfill} & 160 & 0.08 & 0.39 & 4.72 & 8 & 38 \\
%\hbox to 0.9in{EG~And\leaders\hbox to 0.8em{\hss.\hss}\hfill}&&&&&& \\
%\hbox to 0.9in{CD~-283719\leaders\hbox to 0.8em{\hss.\hss}\hfill}&&&&&& \\
%\hbox to 0.9in{Hen 3-461\leaders\hbox to 0.8em{\hss.\hss}\hfill} & 160 & 0.02 & 0.02 & 1.00 & 51 & 51  \\
 \hbox to 0.9in{Hen~3-1591\leaders\hbox to 0.8em{\hss.\hss}\hfill} & 160 & 0.05 & 0.05 & 1.02 & 52 & 53  \\
\hline
\hbox to 0.9in{4~Dra(4050)\leaders\hbox to 0.8em{\hss.\hss}\hfill} & 1600 & 0.02 & 0.14 & 7.27 & 6 & 43 \\
\hbox to 0.9in{4~Dra(4060)\leaders\hbox to 0.8em{\hss.\hss}\hfill}& 1600 & 0.03 & 0.33 & 10.26 & 3 & 32 \\
\hbox to 0.9in{CD-28~3719\leaders\hbox to 0.8em{\hss.\hss}\hfill}& 1600 & 0.01 & 0.01 & 1.08 & 47 & 50 \\
 \hbox to 0.9in{Hen~3-1591\leaders\hbox to 0.8em{\hss.\hss}\hfill} & 1600 & 0.02 & 0.03 & 1.30 & 20 & 26  \\
% \hbox to 0.9in{Hen~3-461\leaders\hbox to 0.8em{\hss.\hss}\hfill} & 1600 & 0.01 & 0.01 & 1.00 & 21 & 21  \\
\hline
%\hbox to 0.9in{4~Dra(4050)\leaders\hbox to 0.8em{\hss.\hss}\hfill} & 3200 & 0.02 & 0.13 & 8.20 & 5 & 42 \\
\hbox to 0.9in{4~Dra(4050)\leaders\hbox to 0.8em{\hss.\hss}\hfill} & 3680 & 0.01 & 0.14 & 9.05 & 4 & 40 \\
\hbox to 0.9in{4~Dra(4060)\leaders\hbox to 0.8em{\hss.\hss}\hfill} & 3680 & 0.02 & 0.32 & 13.71 & 2 & 30 \\
\hbox to 0.9in{CD-28~3719\leaders\hbox to 0.8em{\hss.\hss}\hfill}  & 3680 & 0.01 & 0.01 & 1.24 & 39 & 48 \\
%\hbox to 0.8in{\object{EG And}   \leaders\hbox to 0.4em{\hss.\hss}\hfill} & 3680.0 & 0.01 & 0.01 & 1.18 & 52 & 62  \\
 \hbox to 0.9in{EG~And\leaders\hbox to 0.8em{\hss.\hss}\hfill}& 3680 & 0.01 & 0.01 & 1.18 & 52 & 62  \\
%\hbox to 0.9in{Hen~3-461\leaders\hbox to 0.8em{\hss.\hss}\hfill} & 3680 & 0.00 & 0.00 & 0.98 & 14 & 14   \\
%\hline
%4~Dra(4050) & 7360.0 & 0.01 & 0.10 & 9.25 & 4 & 33 \\
%4~Dra(4060) & 7360.0 & 0.02 & 0.23 & 13.64 & 2 & 23  \\
%EG~And & 7320.0 & 0.01 & 0.01 & 1.33 & 41 & 55  \\
%Hen~3-1591& 7320.0 & 0.01 & 0.01 & 1.25 & 11 & 14  \\
%Hen~3-461 & 7360.0 & 0.00 & 0.00 & 1.00 & 10 & 10  \\
%\hline
%CD-28~3719 & 8000.0 & 0.01 & 0.01 & 1.35 & 26 & 34  \\
%Hen~3-1591 & 8000.0 & 0.01 & 0.02 & 1.51 & 11 & 16  \\
%\hline
%4~Dra(4050) & 14720.0 & 0.01 & 0.10 & 12.35 & 2 & 31 \\
%4~Dra(4060) & 14720.0 & 0.01 & 0.19 & 14.97 & 1 & 20 \\
%4~Dra(4060) & 14720.0 & 0.01 & 0.19 & 14.97 & 1 & 20 \\
%CD-28~3719  & 19200.0 & 0.00 & 0.01 & 1.25 & 20 & 25  \\
%EG~And & 14720.0 & 0.00 & 0.01 & 1.49 & 29 & 43  \\
\enddata
\tablenotetext{a}{We list only those bin sizes for which $s \gtrsim s_{exp}$ }
\end{deluxetable}

%\begin{figure*}[!bp]
%   \includegraphics{/home/natalia/PROYECTOS/EEUU/toall/flick_4dra4050-c.ps}
% \caption{Timing analysis(4050)}
%\end{figure*}
%\begin{figure*}[!bp]
%   \includegraphics{/home/natalia/PROYECTOS/EEUU/toall/flick_4dra4060-a.ps}
 %\caption{Timing analysis(4060)}
%\end{figure*}

\section{Discussion and Conclusions}
\label{sec:concl}

We analyzed the deep, broad-band {\em Suzaku} observations of five symbiotic stars, previously known to be X-ray sources.  The spectra obtained with {\em Suzaku} allow to particularly unveil the origin of the X-ray emission in symbiotics when compared with data obtained with ROSAT, limited by energy coverage, or {\em Swift}, limited by continuous observing time. The high temperatures of the plasma (see Table \ref{tab:models}) strongly suggest that the emission originates in an accretion disk boundary layer ($\delta$-type) instead of a colliding wind region ($\beta$-type).  Of the five sources observed with {\em Suzaku}, only CD~-28~3719 retains its previous classification as $\delta$-type source (see Paper I) derived from {\em Swift} observations.  The proposed WD nature of the compact object in Hen~3-1591 and the temperature of the X-ray emitting plasma suggest that it should now be considered a $\delta$-type source instead of $\gamma$-type source \citep[those symbiotics with neutron stars as compact 
objects; see][]{muerset97}. The likely presence of a new soft component in the X-ray spectrum of Hen~3-461 encourage to propose a $\beta/\delta$-type classification. No X-ray spectral type was proposed for 4~Dra and given the results obtained from the spectral fit, we advocate for a $\delta$-type categorization. Finally, the broadband energy coverage now shows that EG~And should be considered a $\delta$-type instead of $\beta$-type source as originally proposed by \citet{muerset97}.  While the $\delta$ components in the \citeauthor{kennea09} sample were all heavily absorbed, this is not universally true of all $\delta$ components; Hen~3-1591, EG~And and 4~Dra are all lightly absorbed and detected below 1 keV.

When compared with earlier data, all sources show changes in their intrinsic flux and N$_{H}$.  In Paper I we proposed a scenario where the $\delta$-type X-ray emission in symbiotics originates in the accretion disk boundary layer, implying that long term changes in the X-ray flux are mostly due to changes in the accretion rate to the disk while changes in the soft X-rays can also be caused by variations in the amount of absorbing material.  It is still unknown where the absorber is located. The days-to-week time scale changes in N$_H$ observed in the $\delta$-type prototype RT~Cru \citep{luna10, luna07} suggest that the absorber is located close to the WD. However it is unclear if these changes are related and if so, how they are related, with the amount of material entering the accretion disk. In our study we witness that although high flux states encompass high N$_{H}$ in the case of CD~-28-3719 and 4~Dra,  in the case of Hen~3-461, N$_{H}$ is higher now while F$_{X}$ is lower than when observed with {\em 
Swift} (see Table 2 in Paper I). 

%%% Hen 3-461 discussion
Our crude spectral fit indicates that the amount of absorption of the hard X-ray component in Hen~3-461 has increased considerably and possibly a separate soft spectral component has emerged. The intrinsic luminosity has decreased by about 30\% since the {\em Swift} observation in 2010.  Although not contemporaneous UV data are available, GALEX (NUV) observations taken one year before our {\em Suzaku} data indicate that the F$_{UV}$=1.86$\times$10$^{-12}$ erg cm$^{-2}$ s$^{-1}$, then F$_{UV}$/F$_{X} \gtrsim$ 0.7 (we quote the lower limit as the reddening for this source is unknown) and thus the accretion disk boundary layer seems to be still in the optically thin regime.  In Paper I we proposed a scenario where the soft emission of $\beta/\delta$-type objects could be related with a colliding-wind region or jets. We could be witnessing the clear-up of a colliding-wind region. The equivalent width of the (unresolved) Fe K$\alpha$ region of around 400 eV resembles the values found by \citet{mukai07} on the 
well-known jet source with a two-component X-ray spectra, CH~Cyg.

%%% Hen 3-1591 discussion
Our best-fit model for \object{Hen~3-1591} points out to the presence of a white dwarf as the accreting compact object instead of a neutron star, i.e. a white dwarf symbiotic instead of a symbiotic X-ray binary as previously though. Evidence supporting this scenario also comes from the optical spectra which shows many emission lines of a variety of ions, similar to that of a planetary nebula, where the strong UV radiation field from the WD photoionizes the surrounding nebulae. \citet{hedrick04} detected the flickering behavior in the B band data of this object, which suggests that the blue light is from an accretion disk.  Hen~3-1591 belongs to a rare subclass of $d^{\prime}$-type yellow symbiotics, with a dusty IR continuum and they belong to the Galactic disk, so low metallicity is not implicated.  They are interpreted as systems in which the hot component has recently evolved from the AGB to the white dwarf stage.  In this interpretation, the dust is from the mass lost by the AGB star, and the nebulosity 
is in fact the planetary nebula (PNe), and neither is due to the present-day giant (see Jorissen et al. 2005 and references therein).  If we compare with the well-known classical novae GK~Per \citep{bode87} and V458~Vul \citep{wesson08}, both claimed to be a classical nova inside a planetary nebula and if the PNe phase is as short as thought (10-20 kyr; \cite{badenes15}), we might question the white dwarf nature of the accretor which should be still a hot subdwarf. 

Hen~3-1591 also shows the {\em barium syndrome}, i.e. overabundances of $s$-process elements and the presence of single-ionized barium which cannot be explained if the red giant is not part of a binary system \citep{jorissen92, jorissen05}. The now-observed red giant had its photosphere polluted by $s$-processed elements by the previously-AGB companion, that now should be a WD. 
Thus, our high signal-to-noise, broadband {\em Suzaku} spectrum adds more support to the presence of a WD or hot subdwarf. If we use the plasma temperature from the cooling flow model to derive the WD mass, we obtain M$_{WD}$= 0.45 M$_{\odot}$, on the low-side of the WD mass distribution.
%and the R$_{WD}$= 10.3$\times$10$^{8}$ cm.

The high temperature of the plasma strongly suggest that in Hen~3-1591 the X-ray emission arise in an accretion disk boundary layer instead of a colliding wind region. The strong-shock condition implies wind speed's of around 3,000 km s$^{-1}$ for the observed temperatures, and such high speeds outflows or winds had not been detected so far in Hen~3-1591 or any other symbiotic. The lack of UV data does not allow to use the ratio of UV and X-ray fluxes as a proxy for the optical depth of the boundary layer. The decrease of temperature and luminosity (i.e. accretion rate) between ASCA and {\em Suzaku} observations, however, suggest that the optical depth of the X-ray emitting plasma has changed, being higher during {\em Suzaku} observation and thus we might have observed an smaller optically thin portion of the boundary layer.

%Our data were not sensitive enough to detect flickering, a distinctive feature of accretion-powered systems, and the lack of UV 

%as a compact object and the X-ray type should now be $\delta$ in the new classification scheme proposed in \citet{Paper1}.

%Symbiotics are variable sources in all time scales. X-ray short term variability has been studied for most symbiotics ({\bf some ref}), however, only recently there are more than single-epoch X-ray data to glimpse the long term variability. Since the first detection with {\em Swift} in 2010, the X-ray flux from Hen~3-461 dropped to values xxx times lower.  In the case of CD~-28~3719, the flux dropped a factor of two since the {\em Swift} observation. Similar changes in the X-ray flux are observed in 4~Dra, EG~And and Hen~3-1591, by factors of five to ten in months to years. 

%The X-ray emission from EG~And, when observed with ROSAT, was classified as $\beta$-type by \citet{muerset97}, i.e. a system where a colliding wind region is responsible for the observed X-ray spectrum.  We observed EG~And with {\em Suzaku} for $\sim$ 100 ks on 2011, and detected the system for the first time in the 0.3-10 keV energy range, though only 330 photons were detected.

%%%%%EG And Discussion
Based on the high temperature found for the X-ray emitting plasma, we can also rule out colliding winds as the source of X-rays in EG~And. This source was identified as a $\beta$-type symbiotic by \citet{muerset97}. Those authors derived a temperature for the X-ray emitting plasma 1.3$\pm$0.5 keV using ROSAT PSPC data.  This value is much lower than the temperature found from our modeling of the {\it Suzaku} spectrum (5-10 keV) and is much likely due to the lack of sensitivity above 2 keV of the ROSAT spectrum.   
%The ephemeris of EG~And \citep{kolb04,vogel1991} indicate that the WD was near eclipse during the {\em Suzaku} observation.
%If EG~And was observed during the eclipse of the WD with {\em Suzaku}, 
%The ROSAT observation in 1991 took place at orbital phase $\phi$=0.15, also near eclipse.
%\citep[as resulting from the ephemeris from]{kolb04}. 
The velocity implied by the best fit to the {\it Suzaku} spectrum is around a few thousand km s$^{-1}$. This is much faster than the highest velocity line features ($\sim$700 km s$^{-1}$, C IV 1548, 1550\AA  absorption features) found in UV spectra from FUSE and STIS by \cite{crowley08}, and difficult to explain in a low mass white dwarf like the one in EG~And \citep[M$_{WD}\approx$0.4 M$_{\odot}$;][]{kolb04}.
%In similar way for 4~Dra, we can estimate the M$_{WD}$ taking into acount kT$_{max}$ in the cooling flow model and  obtained a mass M$_{WD}$=0.5 M$_{\odot}$ and radius R$_{WD}$=9.8$\times$10$^{8}$ cm. 

Instead, we propose that the X-rays originate in the accretion disk boundary layer around a low mass WD. We estimate the WD mass taking into account the relationship between the $kT$ from the spectral modeling ($kT$=7$_{-2}^{+3}$ keV), kT$\sim$(1/10)$\mu$m$_p$(GM$_{WD}$/R$_{WD}$), where $\mu$ is the mean molecular weight and $m_p$ is the proton mass and the WD mass-radius relationship from \citet{hansen94}, we obtain a mass M$_{WD}\approx$0.5 M$_{\odot}$. 

If the hot-component luminosities reported in the literature (16-400 $L_{\odot}$;\cite{vogel92,kolb04}) are due to an accretion-powered white dwarf, then the implied accretion rates are in the range 10$^{-8}$ to 10$^{-7}$ M$_{\odot}$ yr$^{-1}$ for a 0.4 M$_{\odot}$ WD.  This is squarely in the regime where the boundary layer is expected to be optically thick according to models by \cite{popham95}.  The X-ray luminosity is lower than $L_{hot}$, suggesting that the X-rays are produced in a region that remains optically thin at the outer surface of the mostly optically thick boundary layer. Although we expect X-ray emission due to accretion to be highly variable, EG~And is very faint and we do not detect enough photons to be sensitive to low-level variability ($s_{exp}/average \sim$ 50\%).  

%The X-ray emission observed in EG~And therefore represents an interesting case within the class of X-ray emitting symbiotic stars: accretion powered X-rays with a reasonably soft spectrum and little intrinsic absorption.  A softer spectrum is expected in a lower mass white dwarf, although we may be seeing an even lower temperature in EG~And due to the high UV luminosity of the boundary layer.  Following the arguments laid out in Nelson et al. 2011 for RS~Oph, we find that Compton scattering is very likely the dominant source of cooling in the boundary layer in EG~And.  The X-ray luminosity implies an accretion rate through the optically thin part of the boundary layer of just a few 10$^{-12}$ M$_{\odot}$ yr$^{-1}$.  Cooling by Compton scattering will dominate over radiation for a optically thick boundary layer luminosity $\gtrsim$ 2 $\times$ 10$^{31}$ erg s$^{-1}$.  This condition is clearly met in EG~And.

%%%% 4 Dra discussion

Our {\em Suzaku} observations of 4~Dra, when considered along with previous UV observations \citep{skopal}, indicate that 4~Dra is an accretion-powered symbiotic (i.e., no quasi-steady shell burning on the surface of the WD) in which mass transfer at $\sim$10$^{-8}$ M$_{\odot}$ yr$^{-1}$ produces an optically thick boundary layer around a $\sim$0.6 M$_{\odot}$ WD.  We obtain this rough estimate of $M_{WD}$ from our inference that the boundary layer is optically thick. \cite{popham95} suggest that the accretion rates needed to produce a predominantly optically thick boundary layer around WDs of masses 1.0, 0.8, and 0.6 M$_{\odot}$ are greater than 10$^{-7}$, a few times 10$^{-8}$, and $\sim$10$^{-8}$ M$_{\odot}$ yr$^{-1}$, respectively.  Taking the UV flux at the time of the {\em Suzaku} observation to be comparable to the $\sim$10 L$_{\odot}$ detected by IUE, the BL being optically thick suggests that the WD mass is less than about 0.6 M$_{\odot}$ (for which the IUE luminosity corresponds to accretion rates 
of $\sim$10$^{-8}$ M$_{\odot}$ yr$^{-1}$). In contrast to higher-mass WDs with optically thick boundary layers (such as RS~Oph and SS~Cyg), the small amount of X-ray emitting plasma associated with the optically thick boundary layer around 4~Dra is just as hot as would be expected for an optically thin boundary layer.  

We emphasize that our conclusions about the optical depth of the accretion disk boundary layer and its implications on the WD mass depend on the assumed viscosity parameter, $\alpha$ and other assumptions on the models described by \citet{popham95}.  As noted in Paper I, a 30\% change in $\alpha$ leads to a factor-of-a-few change on the threshold accretion rate.  Allowing for this, the white dwarf in 4~Dra is smaller than about 0.7 M$_{\odot}$, still a low mass.

%and our best guess is that the optical depth and size of the accretion disk boundary layer varies from system to system for some yet unknown reason.

%Comparing the two {\em Suzaku} observations, there is some indication that the amount of absorbing material is correlated with X-ray flux.

%Our models for the X-ray emission of 4~Dra suggest that it likely originates in the accretion disk boundary layer. 
Both fitting approaches discussed in Section \ref{sec:res4dra} yield equally possible scenarios. In the case that both spectra are modeled independently, we can understand that the measured temperatures are different because the optical depth of the emitting region has changed, i.e. during the earliest observation the boundary layer was thicker and the measured temperature and fluxes of the X-ray emitting plasma were lower.  On the other hand, if we assume that during both observations the plasma temperature was the same (the optical depth of the boundary layer and the mass of the WD have not changed between observations) we still obtain acceptable fits with an increase in the amount of absorbing material and the intrinsic luminosity during the last observation, perhaps due to an increase in the accretion rate.  

The orbital solution for 4~Dra was studied by \citet{famaey09}, conducting a radial velocity study of the M giant and found a 1,703$\pm$3 days period, an eccentric orbit ($e$=0.3) and T$_{0}$[MJD]= 53204$\pm$19.  The ROSAT observations thus occurred at orbital phases $\phi$=0.33, 0.15 and 0.61, while {\em Suzakus'} occurred at $\phi^{4050}$=0.23 and $\phi^{4060}$=0.57. If the X-ray absorbing material is tied to the orbital motion of the WD, thus the absorption columns obtained from the fits of the {\em Suzaku} observations should not be very different given than during ObsID 4050 the system was near quadrature while during ObsID 4060 the system was in inferior conjunction, thus from the values in Table \ref{tab:models} and since an undisturbed M giant wind would account for only $\sim$a few times 10$^{21}$ cm$^{2}$ \citep{vandenberg06}, we can conclude that the absorber material is located relatively near the white dwarf and probably intimately connected with the physics of accretion.

\acknowledgments
NEN acknowledges Consejo Nacional de Investigaciones Cient\'ificas y T\'ecnicas, Argentina (CONICET) for the Postdoctoral Fellowship. GJML and NEN acknowledge funding from PIP D-4598/2012 and Cooperaci\'on Internacional \#D2771 from Consejo Nacional de Investigaciones Cient\'ificas y T\'ecnicas, Argentina. KM acknowledges support by NASA through an ADAP grant NNX13AJ13G. JLS acknowledges support by NASA through an ADAP grant NNX15AF19G. 
This research has made use of data obtained from the Suzaku satellite, a collaborative mission between the space agencies of Japan (JAXA) and the USA (NASA) and the VizieR catalogue access tool, CDS, Strasbourg, France. The original description of the VizieR service was published in \citet{vizier00}.


%% in parentheses, after the keyword, but they will not be verified.

{\it Facilities:} \facility{Suzaku}.



%\appendix

\end{document}